# Effect of visual cues and video solutions on conceptual tasks

Tianlong Zu[1], Elise Agra[2], John Hutson[3], Lester C. Loschky[3] and N. Sanjay Rebello[1]

[1]*Department of Physics and Astronomy, Purdue University, 525 Northwestern Ave., West Lafayette, IN 47907*
[2]*Department of Psychology, University of Chicago, 5848 S. University Ave., Chicago, IL 60637*
[3]*Department of Psychological Sciences, Kansas State University, 492 Bluemont Hall, Manhattan, KS 66506*

**Abstract.** Transfer of learning is an important objective of education. However, students usually have difficulties in solving physics transfer tasks even after having solved similar problems previously. We investigated if instruction provided using videos containing detailed explanations of previously solved problems will improve students' performance in tackling near and far transfer tasks. We also investigated whether the combination of visual cues followed by video solutions yields further enhancement of students' performance. N=33 students in an algebra-based physics class participated in an interview containing two problem sets each with one initial task, a training session, and a near and far transfer task. For the training, students either received visual cues only, visual cues and a video or only videos, depending on the condition. We compare students' correctness rate on near and far transfer tasks in the three conditions.

**PACS:** 01.40.Fk

## I. INTRODUCTION

Problem solving is a widely studied area in physics education [1]. It entails transfer of learning which is hard to accomplish [2]. Visual cueing which highlighting the thematically relevant information on a problem diagram can facilitate improvement in students' performance on near transfer problems but not on far transfer problems [3]. Operationally, we define near transfer problems as sharing same underlying principles and representation, but differing in surface features. Far transfer problems share same underlying principles but differ in both surface features and representation. There could be several reasons for the failure of visual cues to facilitate far transfer. One reason is that visual cues provide information in only the visual channel. Mayer's modality principle [4] states that people learn better from information presented through two channels (visual and audio) rather than a single channel. To test this effect, we designed video solutions (animation with narration) that demonstrated how to solve a problem. The readers can find these videos used in our study from this link [7]. We addressed the following research question:

*How does performance on near and far transfer problems change when students are trained using video solutions, visual cues, or their combination?*

## II. THEORETICAL BACKGROUND

Based on the modality principle we hypothesize that providing students with video solutions can produce better performance on transfer tasks than with visual cues alone. But one can argue that video solutions create greater cognitive load and may thus suppress performance. To address this issue we embed the video solutions along with cues in Karplus' learning cycle [5]. We propose providing visual cues in the exploration phase of the learning cycle where students explore materials and ideas with no other help. Then in the concept introduction phase during which students will be introduced new terms or principles by teacher, we provide the video solutions. In this way the presentation of the video solution is scaffolded by the visual cues. This sequence of exploration followed by a more didactic explanation phase is consistent with Bransford's and Schwartz's notion of a "time for telling" in which direct instruction is most effective when it is preceded by a more exploratory phase of learning [6]. In light of these ideas, we generate a second hypothesis that visual cueing followed by video solutions should produce the best learning outcomes.

In summary, we can test two competing hypotheses: video solutions produce best learning (based on modality principle) versus visual cues preceding video solutions produce best learning (based on learning cycle). The research design described below tests these competing hypotheses.

## III. METHOD

N=33 students enrolled in an algebra-based physics class at a large U.S. Midwestern university received extra credit equal to 1% of their course grade as incentive for participation in our study. The materials were conceptual physics problems with diagrams similar to those from a previous study [3]. Students were randomly assigned to three conditions. Each condition solved two problem sets. Each set contained an initial problem, a training session, a near transfer problem, and a far transfer problem. The two problem sets were randomized and the training problems within each set were also randomized to eliminate order effects. The initial, near and far transfer sessions were identical across the three conditions but the training session differed across the conditions as per Table 1.

The training session was designed to ensure equal time-on-task across conditions. We did not have a fourth

(control) condition receiving neither solutions nor cues because previous work has already shown that the control condition is outperformed by the cued condition [3]. Students were interviewed individually. Each interview lasted around 30 minutes and was video and audio recorded. Sample problems used in each problem set are shown in Fig. 1. Also, for the validation of the correctness and appropriateness of the videos, three physics professors and two graduate TAs were asked to inspect the video solutions.

Our goal was to determine the impact of the videos and/or visual cues on the correctness of students' responses. Students' responses were coded as correct if both their answer and reasoning were scientifically correct. Alternatively, students' responses were coded as incorrect if either the answer or the reasoning was scientifically incorrect.

We conducted Fisher's Exact Test, a variant of the Chi-Squared test for small numbers of participants, to investigate the differences between conditions on the initial, near transfer, and far transfer problems for the two problem sets.

**TABLE 1.** The training session in each condition.

| Condition | Training Session |
|---|---|
| Cue + Video | Visual cues on two examples + Video solution of one example |
| Video Only | Video solutions of two examples |
| Cue Only | Visual cues on four examples |

## IV. RESULTS

We investigated the correctness rate of students' responses to the initial, near transfer, and far transfer problems for all three conditions. Figure 2 shows the combined results of both problem sets.

The Fisher's exact test showed no significant differences between groups on the initial problem, $\chi^2 (2, 66) = 0.20$, $p = 1.00$. Thus, participants in the three conditions can be considered as equivalent. For the near transfer problem, we found significant differences between the three groups, $\chi^2 (2, 66) = 9.21$, $p = .008$. Both the Video + Cue and Video Only groups contributed to the significant difference, as determined by the adjusted residuals. Finally, we found no significant differences between the three groups on the far transfer problem, $\chi^2 (2, 66) = 2.5$, $p = .388$.

We also investigated how students' initial correctness moderated their performance on the near and far transfer session. A Fisher's Exact Test was run to compare the students' performance on near and far transfer session in terms of their initial correctness. The results are shown in Figures 3 and 4.

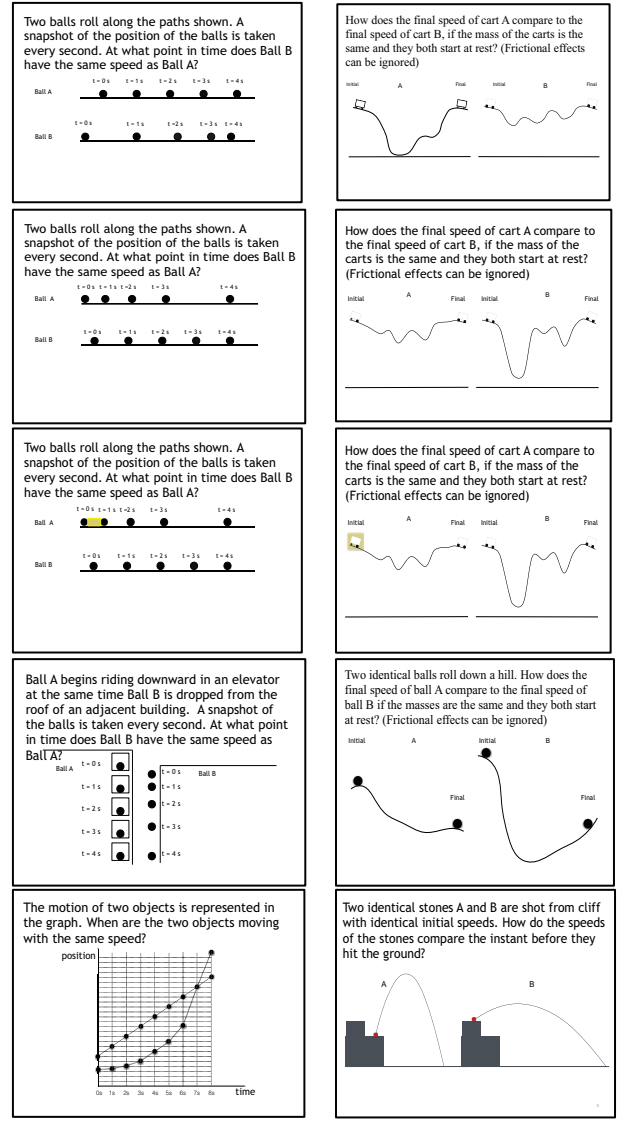

**FIG 1.** Problems used in the Ball problem set (left) and Cart problem set (right). From top to bottom are initial, example training, example training with visual cues added, near transfer and far transfer problem.

For the students who solved the initial task incorrectly, we found a significant difference between groups on the near transfer tasks, $\chi^2 (2, 26) = 6.317$, $p = .013$, but no significant difference on the far transfer task. On the near transfer task, both the Cue + Video and the Video Only conditions outperformed the Cue Only condition.

For students who solved the initial problem correctly, we found a significant difference on both the near ($\chi^2 (2, 26) = 6.317$, $p = .013$) and far transfer tasks $\chi^2 (2, 26) = 7.044$, $p = .039$. In both the near and far transfer tasks, participants from the Cue + Video condition and Video Only condition outperformed those from Cue Only condition.

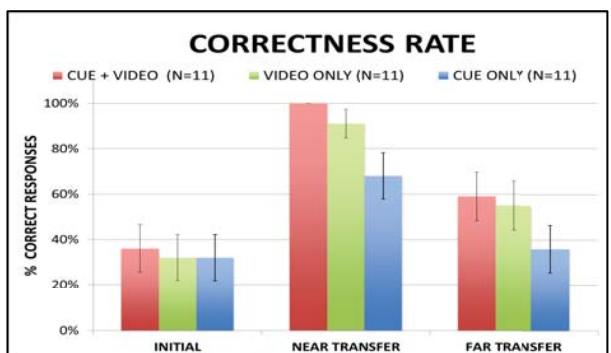

**FIG 2.** Correctness rate across the three conditions for the initial, near transfer, and far transfer problems. Due to the ceiling effect on near transfer problem, there is no error bar.

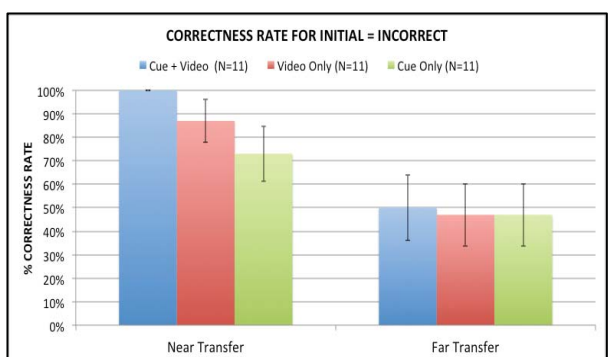

**FIG 3.** For students solving the initial problem incorrectly, correctness rate across the three conditions for the near transfer, and far transfer problems. Due to the ceiling effect on near transfer problem, there is no error bar.

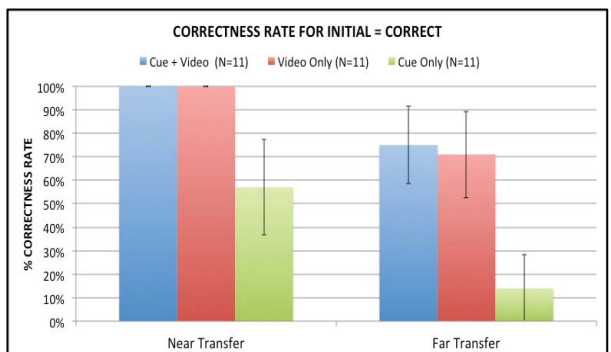

**FIG 4.** For students solving the initial problem correctly, correctness rate across the three conditions for the near transfer, and far transfer problems. Due to the ceiling effect on near transfer problem, there is no error bar.

Taken together, these results seem to suggest that Cue + Video and the Video only conditions outperform the Cue only conditions on the far transfer problems, only if the students possess a certain amount of prior knowledge as evidenced by their correct responses to the initial task.

## V. CONCLUSIONS & DISCUSSION

We investigated the effects of providing visual cues, video solutions, and combinations of the two on performance on transfer problems. We found evidence supporting the first hypothesis that learning from watching video solutions yields better performance than leaning by visual cueing alone in the transfer session. This is consistent with Mayer's modality principle that learning from visuals and narrations together is superior to learning from either visuals or narrations alone.

We also found evidence tending to support the second hypothesis that the use of a combination of visual cues followed by a video solution showed better enhancement of problem solving performance on the near transfer problems than merely providing video solutions or visual cues alone. However, there was no significant difference between these groups on the far transfer problem.

We also investigated how initial correctness moderates students' performance on the transfer problem. Here we found that for the students who could not solve the initial problem correctly, the Cue + Video group significantly outperformed those from the Video Only and Cue Only group. However, this result could be due to the ceiling effect on the near transfer problem, since all of the students in the Cue + Video group solved the near transfer problem correctly. Furthermore, there was no significant difference between different groups on the far transfer session.

For the students who solved initial problems correctly, students from the Cue + Video group and Video Only group significantly outperformed those from the Cue Only group on both near and far transfer problems. Again, we do not know if the significant difference is due to the ceiling effect on the near transfer problem. However we do see that the significant difference holds on the far transfer problems which do not have ceiling effect, this will place some confidence to the result on the near transfer session since far transfer by its definition is harder problems than near transfer. These results seems to suggest that students who possess greater prior knowledge (as evidenced by correct responses to the initial problem) are more likely to be able to use what they have learned from the video solutions or a combination of visual cues and video solutions to correctly solve the transfer problem.

These results suggest that visual cueing in combination with video solutions can promote near transfer, but not necessarily far transfer, especially for students who do not possess the requisite prior knowledge. This is because visual cueing highlights the thematically relevant information thereby facilitating students to attend to the relevant information. The video solutions help students learn the necessary procedural knowledge that would facilitate them to use the thematically relevant information in the problem diagram. If students have adequate prior knowledge, learning from watching videos can help consolidate their knowledge resulting in better transfer performance.

These results are consistent with Mayer's modality principle, which suggests that learning which utilizes both the visual and auditory channels is more effective than learning which utilizes only one of the channels [4]. The video solution is a multimedia presentation which utilizes both channels, while the cues utilize only the visual channel.

These results are somewhat consistent with the learning cycle approach by Karplus and the time for telling approach by Bransford and Schwartz [5, 6]. Both of these strategies suggest that exploration followed by direct instruction is more effective than either pure exploratory learning or learning through direct instruction. We have evidence, albeit not statistically significant, that a combination of visual cues (which are akin to exploration) followed by video solutions (which is akin to direct instruction) produce better performance on transfer problems than either pure exploration using visual cues or pure direct instruction using video solutions.

## VI.   LIMITATIONS & IMPLICATIONS

The main limitation of this study is clearly the small number of participants. We would need to do a larger study with more participants to become more confident of the trends that we see in our results. Another important limitation is that we have conducted this study on only two sets of problems, both of which are in the area of introductory mechanics. Similar studies with a wider range of applicability would be needed to study if the results would be more generally applicable. Finally, another very important limitation pertains to the ecological validity of this study. The study was conducted in a laboratory with students recruited from an introductory physics course. It would be interesting to conduct another study with students who are actually using visual cues and/or video solutions to solve problems as part of their usual coursework. In spite of these limitations, these results have implications in the use of online tutorials often used to facilitate learning and problem solving. They seem to suggest that the use of visual cues or hints that facilitate students to explore solving the problem by themselves before being provided direct instruction on solving the problem might be an effective strategy to promote transfer of learning and problem solving.

## ACKNOWLEDGMENTS

This work is supported in part by the U.S. National Science Foundation grant 1348857.

[1] L. Hsu, et al., *American Journal of Physics* **72**, 9 (2004): 1147-1156.

[2] D. N. Perkins, and G. Salomon, *International Encyclopedia of Education*, **2** (1992).

[3] A. Rouinfar, et al., in Proceedings of the Physics Education Research Conference, Minneapolis, MN, 2014, edited by P. V. Engelhardt, A. D. Churukian, D. L. Jones, (AIP, 2014).

[4] R. Mayer, *The Cambridge Handbook of Multimedia Learning* (Cambridge University Press, New York) **3**, 2001).

[5] R. J. Karplus, *J. Res. Sci. Teach.* **14**, 169 (1977).

[6] D. L. Schwartz, and J. D. Bransford, *Cognition and Instruction* **16**, 4 (1998): 475-5223.

[7] https://www.youtube.com/channel/UCWObNiP3BRt3TDTYR8_JLUQ/videos